\documentclass{article}
\usepackage{graphicx} 
\usepackage{authblk}
\usepackage[colorlinks,linkcolor=blue,citecolor=red,urlcolor=blue]{hyperref}
\usepackage{amsmath,amsfonts,amssymb,amsthm}
\usepackage{nicefrac}  
\usepackage{natbib}
\usepackage[margin=2.5cm]{geometry}

\newcommand{\KL}{\mathop{\mathrm{KL}}\nolimits}
\newcommand{\cP}{\mathcal P}
\newcommand{\cM}{\mathcal M}
\newcommand{\cF}{\mathcal F}
\newcommand{\R}{\mathbb R}
\newcommand{\X}{E}
\newcommand{\m}{\mathtt m}
\newcommand{\grad}{\textrm {grad}}
\newcommand{\ssp}[1]{\langle #1 \rangle}
\newcommand{\norm}[1]{\|#1\|}

\newtheorem{definition}{Definition}
\newtheorem{proposition}{Proposition}
\newtheorem{remark}{Remark}

\usepackage{todonotes}

\title{A note on the unique properties of the Kullback--Leibler divergence for sampling via gradient flows}
\author[1]{Francesca Romana Crucinio\thanks{ \href{mailto:francescaromana.crucinio@unito.it}{francescaromana.crucinio@unito.it}}}
\affil[1]{ESOMAS, University of Turin, Italy \& Collegio Carlo Alberto, Turin, Italy}
\date{ }

\begin{document}

\maketitle

\begin{abstract}

We consider the problem of sampling from a probability distribution $\pi$ which admits a density w.r.t. a dominating measure. It is well known that this can be written as an optimisation problem over the space of probability distributions in which we aim to minimise a divergence from $\pi$. The optimisation problem is normally solved through gradient flows in the space of probability distributions with an appropriate metric. We show that the Kullback--Leibler divergence is the only divergence in the family of Bregman divergences whose gradient flow w.r.t. many popular metrics does not require knowledge of the normalising constant of $\pi$.
\end{abstract}

\section{Introduction}
Sampling from a target probability distribution whose density is known up to a normalisation constant is a fundamental task in computational statistics and machine learning. A natural way to formulate this task is optimisation of a functional measuring the dissimilarity to the target probability distribution.
Following this point of view, one can derive many popular sampling frameworks including variational inference \citep{blei2017variational}, algorithms based on diffusions \citep{roberts1996exponential, durmus2019analysis} and deterministic flows \citep{liu2017stein}, and algorithms based on importance sampling \citep{chopin2023connection, crucinio2025}.
The connection between minimisation of a divergence and Monte Carlo algorithms is established through gradient flows over the space of probability measures (see, e.g., \cite{chewi2025statistical, carrillo_fisher-rao_2024} for a recent review); with different metrics over this space leading to different differential equations whose discretisations correspond to many popular Monte Carlo algorithms.

The most widely used divergence is the reverse Kullback--Leibler (KL) divergence whose gradient flow w.r.t. the Wasserstein-2 metric can be implemented by a Langevin diffusion \citep{jordan1998variational} and easily discretised in time, resulting  in the unadjusted Langevin algorithm \citep{roberts1996exponential}. Recently, gradient flows of the reverse Kullback--Leibler divergence w.r.t. the Fisher--Rao metric have also been considered \citep{Lu2019, crucinio2025, chen2023sampling, domingo-enrich2023an}. The Fisher--Rao gradient flow is intimately tied to mirror descent dynamics \citep{chopin2023connection} and can be efficiently implemented via importance sampling \citep{crucinio2025, korba2022adaptive, dai2016provable}.
Combination of the Wasserstein and Fisher--Rao metric leads to the Wasserstein--Fisher--Rao gradient flow, which, in the case of the Kullback--Leibler divergence, corresponds to a combination of exploration moves using a Langevin diffusion and reweighting \citep{chewi2025statistical, lambert2022variational, yan2024learning}, and can be implemented via sequential Monte Carlo \citep{crucinio2025}.

Alternatives to the previous methods which are based on Markov chains and importance sampling are given by deterministic ODE flows. 
\cite{Nusken2024, maurais2024sampling} propose ODE flows to approximate the Fisher--Rao gradient flow of KL.  \cite{liu2017stein} obtains an ODE to approximate the Wasserstein gradient flow of KL which can be interpreted as a gradient flow of the KL divergence w.r.t. a kernelised Wasserstein metric \citep{duncan2023geometry, chewi2020svgd}.

Despite the widespread use of the KL divergence to derive gradient flows, it is well known that other families of divergences, e.g. $f$-divergences and Bregman divergences, are very flexible and allow to enforce desirable properties such as ability to capture tail behaviour or target the mode \citep{regli2018alpha}.
The rationale behind the choice of KL divergence is that this divergence leads to Monte Carlo sampling algorithms which do not require knowledge of the normalising constant of the target, and can therefore be implemented in general settings.

In this note, we show that the Kullback--Leibler divergence is the only divergence within the class of Bregman divergences which leads to a gradient flow which does not depend on the normalising constant of $\pi$. We consider the case in which $\pi$ admits a density w.r.t. a dominating measure, which covers both measures which admit a density w.r.t. Lebesgue and discrete measures.
Our result (Proposition~\ref{prop:main}) complements that of \citet[Theorem 4.1]{chen2023sampling} which establishes an equivalent result for the class of $f$-divergences in the case of measures which admit a density w.r.t. Lebesgue.

Our result covers the case in which the gradient flow of the chosen divergence remains unaltered if the target is multiplied by a constant $c$. More generally, one can ask when replacing the target $\pi$ with $c\pi$ leads to equivalent minimisation problems (i.e. minimisation problems with the same minimisers). We consider this relaxed requirement in Section~\ref{sec:conditions}.

\subsection*{Notation}
Let $\X$ be a topological vector space endowed with the Borel $\sigma$-field $\mathcal{B}(\X)$. 
We denote by $\cM(\X)$ the vector space of (signed) measures and by $\cM_\lambda(\X)$ the subset of $\cM(\X)$ of measures which admit a density w.r.t. a dominating measure denoted $\lambda$. 
Let $\cM^*(\X)$ be the dual of $\cM(\X)$.
    For $\mu\in \cM(\X)$ and $f\in \cM^*(\X)$, we denote $\ssp{f, \mu}=\int f(x)\mu(dx)$. 
We denote by $\mathcal{P}(\X)$ the set of probability measures over
$\mathcal{B}(\X)$ and endow this space with the topology of weak convergence.  When $\X = \R^d$, the subset of $\mathcal{P}(\R^d)$ of probability measures admitting a density w.r.t. Lebesgue is denoted by $\mathcal{P}^{ac}(\R^d)$. 
For all $p\geq 1$, we denote by
$\mathcal{P}_p(\X) = \{\mu \in \mathcal{P}(\X):\int
  \norm{x}^p d \mu(x) < +\infty\}$ the set of probability measures over
$\mathcal{B}(\X)$ with finite $p$-th moment.   

\section{Background}

\subsection{Sampling as optimisation over measures}
\label{sec:gf}

Let us denote the target distribution over $\X$ by $\pi$ and 
let $\cF:\cM(\X)\rightarrow \R^+$ be a functional on $\cM(\X)$ measuring the distance between $\mu$ and $\pi$. Consider the optimisation problem
\begin{equation}
\label{eq:opt}
    \min_{\mu \in \cP(\X)} \cF(\mu),
\end{equation}
that is the minimum of $\cF$ restricted to the space of probability measures, which recovers $\pi$ if solved exactly.
The optimisation problem~\eqref{eq:opt} can be solved by following the direction of steepest descent of $\cF$, given by the gradient flow partial differential equation (PDE)
\begin{align}
    \label{eq:pde}
    \partial_t\mu_t = -\grad_{\m} \cF(\mu_t),
\end{align}
where $\grad_{\m} \cF(\mu)$ denotes the gradient of $\cF$ w.r.t. metric $\m$ \citep{ambrosio2008gradient,carrillo_fisher-rao_2024}.
Depending on the choice of metric $\m$, the gradient $\grad_{\m}\cF(\mu)$ takes different shapes; for most commonly used metrics the gradient is related to the natural gradient (also known as first variation) of $\cF$ \citep{amari1998natural}:
\begin{definition}\label{def:first_var}
If it exists, the first variation of $\cF$ is the functional $\frac{\delta \cF}{\delta \mu}(\nu):\X \rightarrow \R$ s. t. for any $\mu \in \cM(\X)$, with $\xi = \mu-\nu$:
\begin{equation}
\label{eq:first_var}
\lim_{\epsilon \rightarrow 0}\nicefrac{1}{\epsilon}(\cF(\nu+\epsilon  \xi) -\cF(\nu))
=\ssp{\frac{\delta \cF}{\delta \mu}(\nu), \xi}
\end{equation} 
and is defined uniquely up to an additive constant.
\end{definition}

Restricting the optimisation problem in~\eqref{eq:opt} to the set of probability measures which admit a density w.r.t. the Lebesgue measure and have finite second moment, $\mathcal{P}_2^{ac}(\R^d)$, and considering the
$2$-Wasserstein distance as metric, one obtains the Wasserstein gradient flow \citep{ambrosio2008gradient}
\begin{align*}
    \partial_t \mu_t=  -\grad_{\mathtt{W}}\cF(\mu_t)= \nabla \cdot \left( \mu_t\nabla \frac{\delta \cF}{\delta \mu}(\mu_t) \right)  \quad t \in [0,\infty),
\end{align*}
where $\nabla $ denotes the gradient in $\R^d$.
Wasserstein gradient flows are typically defined over $\mathcal{P}_2^{ac}(\R^d)$, although extensions to discrete spaces exist \citep{sun2023discrete}.

Selecting $\cF(\mu)=\KL(\mu|\pi)$ leads to the Fokker--Planck equation of the Langevin diffusion \citep{jordan1998variational}
\begin{align}
\label{eq:langevin}
    dX_t = \nabla\log\pi(X_t)dt + \sqrt{2}dB_t,
\end{align}
where $(B_t)_{t\geq 0}$ is a $d$-dimensional Brownian motion.
The Euler--Maruyama discretisation of~\eqref{eq:langevin} is known as the unadjusted Langevin algorithm (ULA; \cite{durmus2019analysis}), which leads to the Metropolis-adjusted Langevin algorithm (MALA; \cite{roberts1996exponential}) when combined with a Metropolis--Hastings accept/reject step.

Closely related to the 2-Wasserstein metric is Stein's metric \citep{duncan2023geometry}, in this case the gradient flow PDE for $\mu, \pi\in \mathcal{P}_2^{ac}(\R^d)$ is given by
\begin{align}
    \partial_t \mu_t&=  -\grad_{\mathtt{STEIN}}\cF(\mu_t)\\
    &= \nabla \cdot \left( \mu_t\int k(x, \cdot)\nabla_x \frac{\delta \cF}{\delta \mu}(\mu_t)( x)\mu_t(dx) \right)  \quad t \in [0,\infty),\notag
\end{align}
where $k$ is a positive semi-definite kernel. In the case $\cF(\mu)=\KL(\mu|\pi)$, the gradient flow w.r.t. Stein's metric corresponds to the continuity equation of Stein variational gradient descent (SVGD; \cite{liu2017stein})
\begin{align}
\label{eq:svgd}
    dX_t^i = \frac{1}{N}\sum_{j=1}^N [\nabla\log\pi(X_t^j)k(X_t^i, X_t^j)+\nabla_1 k(X_t^j, X_t^i)]dt,
\end{align}
for $i=1, \dots, N$, where $\nabla_1$ denotes the gradient w.r.t. the first component.

The Fisher--Rao gradient flow, obtained by considering the Fisher--Rao metric, can be defined more generally, for instance for target densities which do not have finite second moment, and for discrete probability distributions.
The Fisher--Rao gradient flow PDE~\eqref{eq:pde} is \citep{carrillo_fisher-rao_2024}
 \begin{align*}
        \partial_t \mu_t &=-\grad_{\mathtt{FR}}\cF(\mu_t)\\
        &= \mu_t \left( -\frac{\delta \cF}{\delta \mu}(\mu_t)+ \mathbb{E}_{\mu_t} \left[  \frac{\delta \cF}{\delta \mu}(\mu_t)\right] \right) \quad t \in [0,\infty).
\end{align*}
Selecting $\cF(\mu)=\KL(\mu|\pi)$ and observing that in this case $\frac{\delta \cF}{\delta \mu}(\mu) = \log (\mu/\pi)$, one obtains the PDE
 \begin{align*}
        \partial_t \mu_t = \mu_t \left( \log \left( \frac{\pi}{\mu_t} \right) - \mathbb{E}_{\mu_t} \left[  \log \left( \frac{\pi}{\mu_t} \right) \right] \right) \quad t \in [0,\infty).
\end{align*} 
This PDE is exactly solved by $\mu_t \propto \pi^{1 - e^{-t}} \mu_0^{e^{-t}}$ \cite[Theorem 4.1]{chen2023sampling} which corresponds to geometric tempering or annealing dynamics \citep{domingo-enrich2023an, chopin2023connection}.
A natural way to numerically approximate the Fisher--Rao flow of $\KL$ is to first discretise time and then use importance sampling \citep{crucinio2025, korba2022adaptive}: given $\mu_t$, one can obtain an approximation of $\mu_{t+1}$ using $\mu_t$ as proposal and weights 
\begin{align*}
    w_{t+1} = \left( \frac{\pi}{\mu_t} \right)^{1-e^{-\gamma}},
\end{align*}
where $\gamma$ denotes the time discretisation step.

When working in $\mathcal{P}_2^{ac}(\R^d)$, combining the Wasserstein and Fisher--Rao metrics one obtains the Wasserstein--Fisher--Rao (WFR) gradient $\grad_{\mathtt{WFR}} \cF(\mu) = \grad_{\mathtt{W}} \cF(\mu)+\grad_{\mathtt{FR}} \cF(\mu)$ \citep{gallouet2017jko}. If $\cF(\mu)=\KL(\mu|\pi)$ the WFR gradient flow is given by the sum of the Wasserstein and of the Fisher--Rao PDE. Numerical approximations can be obtained combining the unadjusted Langevin algorithm with importance sampling which naturally leads to sequential Monte Carlo methods \citep{del2006sequential} as described in \cite{crucinio2025}. Another alternative to combine the Wasserstein and Fisher--Rao evolution is to considered tempered dynamics, as shown in \cite{Chehab2024, crucinio2026properties}.

\subsection{Bregman divergences}
We now introduce Bregman divergences \citep{BREGMAN1967200} a class of divergences between measures which encompasses many commonly used measures of distance. 
We focus on measures which admit a density w.r.t. a dominating measure denoted $\lambda$, a set we denote by $\cM_{\lambda}(\X)$. This encompasses both the case of measures which admit a density w.r.t. the Lebesgue measure, and the case of discrete probability measures, which admit a density w.r.t. the counting measure.

Since we consider measures which admit a density, we consider the following definition of Bregman divergence \citep{cichocki2010families}.

\begin{definition}
Let $\Phi:\R\to \R$ be a convex functional on $\R$. The $\Phi$-Bregman divergence is defined for any $\mu, \pi \in \cM_\lambda(\X)$ by:
\begin{align}
\label{eq:breg_probas}
     B_{\Phi}(\mu|\pi)&=\int [\Phi(\mu(x))-\Phi(\pi(x))]d\lambda(x)\\
    &-\int (\mu(x)-\pi(x))\Phi'(\pi(x))d\lambda(x).\notag
\end{align}
\end{definition}

Setting $\cF(\mu) = B_{\Phi}(\mu|\pi)$ results in $\frac{\delta \cF}{\delta \mu}(\mu)= \Phi'(\mu(x)) - \Phi'(\pi(x))$ as shown in Appendix~\ref{app:first_variation}, where we also denote by $\mu, \pi$ the densities w.r.t. the dominating measure $\lambda$ with a slight abuse of notation.

Among the class of Bregman divergences we recall the Kullback--Leibler divergence 
$\KL(\mu|\pi) =\int \log(\nicefrac{d\mu}{d\pi}) \mu$ obtained considering 
\begin{equation*}
    \Phi(t)= t \log t  -t +1.
\end{equation*}
The Kullback--Leibler divergence is the only Bregman divergence which is also an $f$-divergence \citep{amari2009alpha}.

More generally, 
we can consider $\beta$-divergences \citep{cichocki2010families}, which are obtained setting
\begin{align*}
    \Phi(t) =\begin{cases}
         t\log t- t+1\qquad\qquad\ \textrm{if } \beta=1\\
        t-\log t-1\qquad\qquad\quad \textrm{if } \beta=0\\
        \frac{\beta-1+t^\beta-\beta t}{\beta(\beta-1)}\qquad\qquad \qquad\textrm{otherwise}
    \end{cases}.
\end{align*}
This family encompasses the $L^2$ norm $\nicefrac{1}{2}\|\mu-\pi\|_2^2 = \nicefrac{1}{2}\int |\mu(x)-\pi(x)|^2dx$, obtained for $\beta = 2$ and the Kullback--Leibler divergence obtained for $\beta = 1$.



\begin{remark}
    It is possible to define a wider class of Bregman divergences $B_\phi$ which does not require existence of a density w.r.t. a dominating measure. Given  $\phi:\cM(\X)\rightarrow \R$ a convex functional on $\cM(\X)$. The $\phi$-Bregman divergence is defined for any $\mu, \pi \in \cM(\X)$ by:
\begin{equation*}
     B_{\phi}(\mu|\pi)=\phi(\mu)-\phi(\pi)-\ssp{\frac{\delta \phi}{\delta \mu}(\pi), \mu- \pi}
\end{equation*}
where $\frac{\delta \phi}{\delta \mu}(\pi)$ is the first variation of $\phi$ at $\pi$.
An example of this class is the maximum mean discrepancy (MMD) \citep{JMLR:v13:gretton12a} which can be obtained for $\phi(\mu) = -\int\int k(x,y)d\mu(x)d\mu(y)$  for some fixed positive semi-definite kernel $k$:
\begin{align*}
    \textrm{MMD}^2(\mu, \pi) 
    &=\int\int k(x,y)d\mu(x)d\mu(y)\\
    &+\int\int k(x,y)d\pi(x)d\pi(y)\\
    &-2\int\int k(x,y)d\pi(x)d\mu(y).
\end{align*}
However, the gradient flows of MMD require knowledge of the normalising constant of $\pi$ as noted in \cite{arbel2019maximum}.
\end{remark}

\section{Main Result}

We are now ready to derive our main result which complements \citet[Theorem 3.1]{chen2023sampling} and shows that the reverse Kullback--Leibler divergence is the unique divergence within both the $f$-divergence class and the Bregman divergences class whose gradient flows do not depend on the normalising constant of the target $\pi$.

\begin{proposition}
\label{prop:main}
 Let $\cM_{\lambda}(\X)$ denote the set of measures over $\mathcal{B}(E)$ which admit a density w.r.t. $\lambda$.
    The Kullback--Leibler divergence is the only Bregman divergence such that $B_\Phi(\mu|c\pi) -B_\Phi(\mu|\pi) $ is independent of $\mu\in\cM_{\lambda}(\X)$ for any $c\in(0, +\infty)$ and for any $\pi\in\cM_{\lambda}(\X)$.
\end{proposition}
\begin{proof}
We have that
\begin{align}
\label{eq:proof1}
   B_\Phi(\mu|c\pi) -B_\Phi(\mu|\pi) &= \int \Phi(\pi(x))- \Phi(c\pi(x))d\lambda(x) \\
   &-\int (\mu(x)-c\pi(x))\Phi'(c\pi(x))d\lambda(x) \notag\\
   &+\int (\mu(x)-\pi(x))\Phi'(\pi(x))d\lambda(x), \notag
\end{align}
for any $c\in(0, +\infty)$.

In the case of $\Phi(t) = t\log t-t+1$, corresponding to the Kullback--Leibler divergence, we have
\begin{align*}
   B_\Phi(\mu|c\pi) -B_\Phi(\mu|\pi) = c-1-\log c,
\end{align*}
for any $c\in(0, +\infty)$, showing that the KL satisfies the result.

To establish uniqueness, we follow the proof of \citet[Theorem 3.1]{chen2023sampling} and consider, for $\xi\in\cM_\lambda(\X)$ which integrates to 0, the difference
\begin{align*}
     &[B_\Phi(\mu+\xi|c\pi) -B_\Phi(\mu+\xi|\pi)] - [B_\Phi(\mu|c\pi) -B_\Phi(\mu|\pi)]\\
     &\qquad =  \int [\Phi'(\pi(x))-\Phi'(c\pi(x))]\xi(x) d\lambda(x),
\end{align*}
where we used~\eqref{eq:proof1}.
If $B_\Phi$ is such that $B_\Phi(\mu|c\pi) -B_\Phi(\mu|\pi) $ is independent of $\mu$, then
\begin{align*}
    0&=\int [\Phi'(\pi(x))-\Phi'(c\pi(x))]\xi(x) d\lambda(x).
\end{align*}
Let $f(x) = \Phi'(\pi(x))-\Phi'(c\pi(x))$. The equality
\begin{align*}
    0&= \int f(x)\xi(x)d\lambda(x)
\end{align*}
holds true for all $\xi\in\cM_\lambda(\X)$ which integrate to 0. In particular, it holds true for $\xi(x)d\lambda(x) = \frac{1_A(x)}{\lambda(A)} - \frac{1_B(x)}{\lambda(B)}$ for arbitrary $A, B \in \mathcal{B}(\X)$, where $1_A$ is the indicator function of set $A$. This implies
\begin{align*}
    0 = \frac{1}{\lambda(A)}\int_A f(x) d\lambda(x)
- \frac{1}{\lambda(B)}\int_B f(x) d\lambda(x).
\end{align*}
Thus, 
\begin{align*}
\frac{1}{\lambda(A)}\int_A f(x) d\lambda(x)
= \frac{1}{\lambda(B)}\int_B f(x) d\lambda(x)=: K,
\end{align*}
for all $A, B \in \mathcal{B}(\X)$. 
To see that $f$ is necessarily constant and equal to $K$ almost everywhere consider the set $D = \{x\in \X : f(x) > K + \varepsilon\}$. If $\lambda(D) > 0$, then
\[
\frac{1}{\lambda(D)}\int_{D} f(x) d\lambda(x) > K + \varepsilon,
\]
contradicting the previous identity. Hence $\lambda(D) = 0$. Similarly,
$\lambda(\{x: f(x) < K - \varepsilon\})=0$. Therefore $f(x)=K$ almost everywhere.
Therefore, for almost all  $x\in\X$,
\begin{align}
\label{eq:proof2}
    \Phi'(\pi(x))-\Phi'(c\pi(x)) = K(c),
\end{align}
for some constant $K(c)\in\R$ which depends on $c$.


Let us denote $h(z):=\Phi'(e^z)$ and let $t=\log c$. Then using~\eqref{eq:proof2} we have
\begin{align*}
    h(z+t) &= h(z+\log c) \\
    &= \Phi'(c e^z) \\
    &=\Phi'(e^z)- K(c) \\
    &= h(z)-K(c) \\
    &= h(z)-K(e^t).
\end{align*}
Thus $h(z+t) = h(z)-K(e^t)$ and therefore satisfies \citet[Eq. (2)]{8b1c4492-9d75-38af-ba80-93b1f0ce1772}.

Since $\Phi$ is a convex function, $\Phi'$ is monotone and therefore has a countable number of discontinuities. Thus $\Phi'$ is continuous at at least one point. As $h$ is the composition of $\Phi'$ with $e^z$, we can conclude that $h$ is also continuous at least one point.
By \citet[Theorem 2]{8b1c4492-9d75-38af-ba80-93b1f0ce1772} we can conclude that 
\begin{align*}
    h(z) = z(h(1)-h(0))+h(0) = az+b
\end{align*}
for some $a, b\in\R$ and thus $h$ is an affine function. 
Therefore $\Phi'(t) = h(\log t) = a\log t+b$.



It follows that~\eqref{eq:proof1} is constant iff $$ \Phi(t ) = a t \log t+(b-a)t+C'$$ with $a, b, C'$ constants.
Plugging this $\Phi$ into $B_\Phi$ we obtain
\begin{align*}
    B_\Phi(\mu|\pi) = a\KL(\mu|\pi),
\end{align*}
which completes the proof.
\end{proof}

As a consequence of Proposition~\ref{prop:main}, for every Bregman divergence, $\cF(\mu) = B_{\Phi}(\mu|\pi)$, which is not the Kullback--Leibler divergence, the first variation $\frac{\delta \cF}{\delta \mu}(\mu)= \Phi'(\mu(x))-\Phi'(\pi(x))$ does not remain unchanged if we scale $\pi$
by a positive constant. It follows that the gradient flows discussed in Section~\ref{sec:gf} necessitate knowledge of the normalising constant of $\pi$ to be implemented as replacing $\pi(x)=\gamma(x)/Z$ with its unnormalised version $\gamma$ will lead to a different gradient flow.




\begin{remark}

In general, if one has a divergence $D(\mu|\pi)$ such that 
\begin{align}
    \label{eq:condition}
    D(\mu|\pi)-D(\mu|c\pi)\qquad\textrm{does not depend on $\mu$,}
\end{align}
then the minimiser of $D(\mu|\pi)$ and $D(\mu|c\pi)$ coincide, allowing us to replace $\pi$ with its unnormalised version without altering the resulting minimiser. In addition, under~\eqref{eq:condition} we have $\frac{\delta D(\cdot|\pi)}{\delta \mu}=\frac{\delta D(\cdot|c\pi)}{\delta \mu}$ showing that for divergences satisfying~\eqref{eq:condition} the gradient flow of  $D(\mu|\pi)$ coincides with that of $D(\mu|c\pi)$.

Proposition~\ref{prop:main} and \citet[Theorem 3.1]{chen2023sampling} show that the Kullback--Leibler is the only divergence in the family of Bregman divergences as defined in~\eqref{eq:breg_probas} and in the family of $f$-divergences such that this condition is satisfied. 
As a consequence, the gradient flows in Section~\ref{sec:gf} can be implemented without knowledge of the normalising constant of $\pi$.

There are however divergences outside these families which allow derivation of (Wasserstein) gradient flows without knowledge of the normalising constant of $\pi$. For targets in $\cP_2(\R^d)$, an example is the kernel Stein discrepancy (e.g. \cite{pmlr-v70-gorham17a})
\begin{align*}
    \textrm{KSD}^2(\mu, \pi) &= \int\int k_\pi(x,y)d\mu(x)d\mu(y),
\end{align*}
a special case of MMD which employs the Stein kernel
\begin{align*}
k_\pi(x, y)  &= (\nabla \log \pi(x))^\top k(x,y)(\nabla \log \pi(y))\\
&
+ (\nabla \log \pi(x))^\top \nabla_2 k(x,y)\\
&+ (\nabla \log \pi(y))^\top \nabla_1 k(x,y)
+ \operatorname{tr}(\nabla_1 \nabla_2 k(x,y)).
\end{align*}
As KSD itself does not depend on the normalising constant of $\pi$, also its gradient flows do not depend on this normalising constant.
The Wasserstein gradient flow of this discrepancy has been studied in \cite{korba2021kernel}.

\end{remark}

\section{Alternative Conditions}
\label{sec:conditions}
The condition~\eqref{eq:condition} guarantees that the minimisers of $ D(\mu|\pi)$ and $D(\mu|c\pi)$ coincide and that the first variations $\frac{\delta D(\cdot|\pi)}{\delta \mu}$, $\frac{\delta D(\cdot|c\pi)}{\delta \mu}$ coincide too.

Equality of minimisers can be obtained under the relaxed condition
\begin{align}
\label{eq:relaxed}
    D(\mu|\pi)-D(\mu|c\pi) = K\cdot D(\mu|\pi)+K'
\end{align}
where $K, K'$ do not depend on $\mu$. In this case, we have $D(\mu|c\pi)=(1-K)D(\mu|\pi)+K'$ and thus the minimisers of $D(\mu|\pi)$ and $D(\mu|c\pi)$ coincide.
However, under condition~\eqref{eq:relaxed} the first variation of $D(\mu|\pi)$ and $D(\mu|c\pi)$i thi differ because of the multiplicative constant $1-K$.

We now analyse what~\eqref{eq:relaxed} implies in the Bregman and $f$-divergence families.

\begin{proposition}
    Let  $\phi:\cM(\X)\rightarrow \R$ a be a convex functional on $\cM(\X)$. Consider the $\phi$-Bregman divergence defined for any $\mu, \pi \in \cM(\X)$ by:
\begin{equation*}
     B_{\phi}(\mu|\pi)=\phi(\mu)-\phi(\pi)-\ssp{\frac{\delta \phi}{\delta \mu}(\pi), \mu- \pi}
\end{equation*}
where $\frac{\delta \phi}{\delta \mu}(\pi)$ is the first variation of $\phi$ at $\pi$.
There is no non-trivial divergence $B_{\phi}$ satisfying~\eqref{eq:relaxed}.
\end{proposition}
\begin{proof}
    The condition $$B_\phi(\mu| \pi) - B_\phi(\mu| c\pi) = K \cdot B_\phi(\mu|\pi)+K'$$ implies
\begin{align*}
    B_\phi(\mu|\pi)& - B_\phi(\mu|c\pi) \\
    &= \phi(c\pi)-\phi(\pi) \\
   &+\ssp{\frac{\delta \phi}{\delta \mu}(c\pi), \mu-c\pi} \\
   &-\ssp{\frac{\delta \phi}{\delta \mu}(\pi), \mu-\pi}\\ &= K \left(\phi(\mu)-\phi(\pi)-\ssp{\frac{\delta \phi}{\delta \mu}(\pi), \mu- \pi}\right)+K'\\
   &=K\cdot B_{\phi}(\mu|\pi) +K'
\end{align*}
for all $\mu$ and with fixed $\pi$. This implies 
\begin{align*}
  \ssp{\frac{\delta \phi}{\delta \mu}(c\pi)-\frac{\delta \phi}{\delta \mu}(\pi), \mu} = K\cdot \left(\phi(\mu)-\ssp{\frac{\delta \phi}{\delta \mu}(\pi), \mu}\right)+C
\end{align*}
where $C$ is a constant collecting all terms not depending on $\mu$. For this to be true for all $\mu$ with $K\neq 0$, $\phi$ needs to be an affine function since the l.h.s. is a linear function of $\mu$. Taking $\phi(\mu) = a\int d\mu+b$ results in a degenerate Bregman divergence as $B_{\phi}(\mu|\pi) = 0$ for all $\mu, \pi$. This ends the proof.
\end{proof}

The previous result shows that no non-trivial Bregman divergence  satisfies~\eqref{eq:relaxed}. The next result identifies $f$-divergences that do satisfy~\eqref{eq:relaxed}.

\begin{proposition}
\label{prop:fdiv}
Let $f:
\R^+\to \R$ be a convex function such that $f(1)=0$. Consider the $f$-divergence
$D_f(\mu | \pi)= \int f\left(\frac{d\mu}{d\pi}\right) d\pi$.
The only $f$-divergences satisfying~\eqref{eq:relaxed} are
\begin{itemize}
    \item the forward Kullback--Leibler divergence $\KL(\pi|\mu)$ obtained with $f(s) = \log s$;
    \item $f$-divergences obtained from 
    \begin{align*}
    f(s)= \frac{K'}{1-K-c} + s^{1-\log (1-K)/\log c} p(\log s)
\end{align*}
with $p$ a $T$-periodic function such that $p(0) = -\frac{K'}{1-K-c}$ and
\begin{align}
\label{eq:p_condition}
&p(\ell x+(1-\ell)y)\\
&\qquad\le
\ell r^{\frac{(1-\ell)(x-y)}{T}}p(x)
+
(1-\ell)r^{-\frac{\ell(x-y)}{T}}p(y)\notag
\end{align}
with $r=c/(1-K)>0$ for all $x, y\in\R$ and $\ell\in[0, 1]$.
\end{itemize}
\end{proposition}
\begin{proof}


Condition~\eqref{eq:relaxed} implies
\begin{align*}
D_f(\mu | \pi) &- D_f(\mu | c\pi)\\
&= \int f\left(\frac{d\mu}{d\pi}\right)d\pi
- \int c f\left(\frac{1}{c}\frac{d\mu}{d\pi}\right) d\pi \\
&= \int \left[
f\left(\frac{d\mu}{d\pi}\right)
- c f\left(\frac{1}{c}\frac{d\mu}{d\pi}\right)
\right] d\pi\\
&= K\cdot \int f\left(\frac{d\mu}{d\pi}\right)d\pi +K'\\
&=K \cdot D_f(\mu |\pi) + K'.
\end{align*}
As $\pi\in\cP(\X)$, this can only be satisfied if
\begin{align*}
    f\left(\frac{d\mu}{d\pi}(x)\right)-c f\left(\frac{1}{c}\frac{d\mu}{d\pi}(x)\right) = K\cdot f\left(\frac{d\mu}{d\pi}(x)\right)+K',
\end{align*}
for all $x\in \X$, i.e. $f$ satisfies the functional equality
\begin{align}
\label{eq:functional_eq}
    (1-K)f(s) 
= c f\!\left(\frac{s}{c}\right)+K', \quad \forall s>0.
\end{align}

\paragraph*{Case $c=1-K$.}
In Appendix~\ref{sec:functional_eq} we show that if $c=1-K$ the only solution of~\eqref{eq:functional_eq} which is convex and satisfies $f(1)=0$ is
\begin{align*}
    f(s)=\frac{K'}{(1-K) \log c}\log s \propto \log s
\end{align*}
which gives
\begin{align*}
    D_f(\mu|\pi) \propto \KL(\pi|\mu).
\end{align*}

\paragraph*{Case $c\neq 1-K$.}
In Appendix~\ref{sec:functional_eq} we show that if $c\neq 1-K$ the solutions of~\eqref{eq:functional_eq} which are convex and satisfy $f(1)=0$ are of the form 
\begin{align*}
    f(s)= \frac{K'}{1-K-c} + s^{1-\log (1-K)/\log c} p(\log s)
\end{align*}
with $p$ a $T$-periodic function such that $p(0) = -\frac{K'}{1-K-c}$ satisfying~\eqref{eq:p_condition}.

\end{proof}

The $f$-divergences identified by Proposition~\ref{prop:fdiv} include the forward KL divergence obtained for $f(s) = \log s$. Gradient flows of this divergence have been studied in \cite{zhu2025inclusive}.

Another special case of the $f$-divergences identified by Proposition~\ref{prop:fdiv} is obtained when $p(t)$ is constant. In this case $f$ takes the form
\begin{align*}
    f(s) =  a(s^{\gamma}-1)
\end{align*}
for $\gamma\geq 1$. The resulting $f$-divergences
\begin{align*}
    D_f(\mu|\pi) = a\left(\int \mu(x)^\gamma \pi(x)^{1-\gamma}d\lambda(x) -1\right)
\end{align*}
for $\mu, \pi$ admitting densities w.r.t. $\lambda$, are proportional to power divergences.
In particular for $\gamma=2$ we recover the $\chi^2$ divergence.
Gradient flows of the $\chi^2$ divergence or closely related divergences do exist, e.g. \cite{chewi2020svgd} shows that the SVGD dynamics~\eqref{eq:svgd} can be interpreted as a kernelised Wasserstein gradient flow of the $\chi^2$ divergence, \cite[Section 3]{lu2023birth} derive a Fisher--Rao gradient flow for $\chi^2$ and \cite{li2023sampling} considers a mollified version of $\chi^2$ to derive a deterministic flow.

\section{Discussion}

Proposition~\ref{prop:main} shows that the Kullback--Leibler divergence is the only divergence within the Bregman class defined in~\eqref{eq:breg_probas} whose gradient flows do not require knowledge of the normalising constant of $\pi$. Combining this result with \citet[Theorem 4.1]{chen2023sampling} reduces the class of divergences for which one can develop sampling algorithms based on gradient flows which do not require knowledge of the normalising constant. In particular, our result shows that $\alpha$--divergences and $\beta$--divergences lead to gradient flows requiring the normalising constant.

Another option to build gradient flows that do not require knowledge of the normalising constant of $\pi$ is to consider divergences which satisfy~\eqref{eq:relaxed}. In Section~\ref{sec:conditions} we show that there exist $f$-divergences which satisfy~\eqref{eq:relaxed} but unfortunately their gradient flows are notoriously hard to implement without additional approximations.

Our result and \citet[Theorem 4.1]{chen2023sampling} do not cover the class of kernel discrepancies, which with a careful choice of the kernel can lead to gradient flows that do not require knowledge of the normalising constant of the target \citep{korba2021kernel}. In some cases, kernel discrepancies also allow to obtain gradient flow dynamics which can be implemented if a sample from the target $\pi$ is already available \cite{arbel2019maximum, belhadji2025weighted}.


\section*{Acknowledgments} 
FRC gratefully acknowledges the ``de Castro" Statistics Initative at the \textit{Collegio Carlo Alberto} and the \textit{Fondazione Franca e Diego de Castro}.
FRC is supported by the Gruppo
Nazionale per l'Analisi Matematica, la Probabilità e le loro Applicazioni (GNAMPA-INdAM).

\bibliographystyle{plainnat}
\bibliography{biblio}

\appendix
\section{First variation of Bregman divergence}
\label{app:first_variation}

Let
\begin{align*}
\cF(\nu)&=\int [\Phi(\nu(x))-\Phi(\pi(x))]d\lambda(x)\\
    &-\int (\nu(x)-\pi(x))\Phi'(\pi(x))d\lambda(x).
\end{align*}
Take $\epsilon>0$ and $\xi = \mu-\nu$ a measure which integrates to 0. Then
\begin{align*}
    &\frac{\cF(\nu+\epsilon  \xi) -\cF(\nu)}{\epsilon}\\
    &= \int \frac{\Phi(\nu(x)+\epsilon  \xi(x))-\Phi(\nu(x))}{\epsilon}d\lambda(x)\\
    &-\int  \xi(x)\Phi'(\pi(x))d\lambda(x).
\end{align*}
Taking the limit $\epsilon\to 0$ we find
\begin{align*}
    &\lim_{\epsilon\to 0}\frac{\cF(\nu+\epsilon  \xi) -\cF(\nu)}{\epsilon}\\
    &= \int  \xi(x)[\Phi'(\nu(x))-\Phi'(\pi(x))]d\lambda(x),
\end{align*}
showing that 
\begin{align*}
    \frac{\delta \cF}{\delta \nu}(\nu)= \Phi'(\nu(x)) - \Phi'(\pi(x)).
\end{align*}

\section{Solution of $f$-divergence functional equation}
\label{sec:functional_eq}
Use the substitution $s=e^t$ and $g(t)=f(e^t)$ to write~\eqref{eq:functional_eq} as
\begin{align*}
(1-K) g(t) - cg(t-\log c) = K'.
\end{align*}

Let $T=\log c$, $C:=\frac{K'}{1-K}$, $r:=c/(1-K)$ and
consider the functional equation
\begin{align*}
g(t)=r g(t-T)+C
\end{align*}
for all $t\in \R$.
Iterating
for $n$ steps we find
\begin{align}
\label{eq:nonhomogeneous}
g(t)&=r^n g(t-nT)+C\sum_{k=0}^{n-1} r^k.
\end{align}
\paragraph*{Case $r=1$.}
For the case $r=1$, \eqref{eq:nonhomogeneous} becomes
\begin{align}
\label{eq:base2}
  g(t)&=g(t-nT)+nC.
\end{align}
Fix $t\in\R$ and write 
$t=s+nT$ for $s\in[0,T)$ and $n\in\mathbb{Z}$. 
Then $g(t)=g(s)+nC$.
Since \(n=\frac{t-s}{T}\), we obtain
\[
g(t)=g(s)+\frac{C}{T}(t-s)
=\frac{C}{T}t+\left(g(s)-\frac{C}{T}s\right).
\]
Define $p(t):=g(t)-\frac{C}{T}t$, then
\begin{align*}
   p(t+T)&=g(t+T)-\frac{C}{T}(t+T)\\
   &=g(t)+C-\frac{C}{T}(t+T)\\
   &=g(t)-\frac{C}{T}t\\
&= p(t)
\end{align*}
where we used $g(t+T)=g(t)+C$ following~\eqref{eq:base2}. Thus \(p\) is \(T\)-periodic, and
\[
g(t)=\frac{C}{T}t+p(t),
\quad p(t+T)=p(t).
\]

We now impose the conditions $f(1)=0$ and $f$ convex to check if a suitable $f$ exists.
Since \(t \mapsto e^t\) is an increasing convex function, it is sufficient to check that $g$ is convex in $\R$ to guarantee that $f$ is convex on $(0,\infty)$.
Since $g(t)$ is the sum of a linear term and $p(t)$, it is convex if $p$ is convex.
However, a periodic function can be convex only if it is constant.
Using \(f(1)=0\) we find $p(0)=0$ and thus  \(p\equiv 0\), and
\begin{align*}
    f(s)=\frac{C}{ T}\log s
\end{align*}

\paragraph*{Case $r\neq 1, r>0$.}
If $r\neq 1$ \eqref{eq:nonhomogeneous} becomes
\begin{align*}
g(t)&=r^n g(t-nT)+C\frac{1-r^n}{1-r}.
\end{align*}

Let $A:=\frac{C}{1-r}$. Then the identity becomes
\begin{align}
\label{eq:base1}
g(t)-A=r^n [g(t-nT)-A]
\end{align}
for all $n\in\mathbb{N}$.
Define $p(t):= r^{-t/T}(g(t)-A)$.
Then, using~\eqref{eq:base1} with $n=1$ we find
\begin{align*}
    p(t) &= r^{-t/T}(g(t)-A)\\
&= r^{-t/T+1} (g(t-T)-A)\\
&= r^{-(t-T)/T}(g(t-T)-A)\\
&= p(t-T).
\end{align*}
Hence,  \(p\) is \(T\)-periodic.
Finally, solving for \(g\),
\[
g(t)= \frac{C}{1-r} + r^{t/T} p(t),
\quad p(t+T)=p(t).
\]
Thus the candidate $f$ is
\begin{align*}
    f(s)= \frac{C}{1-r} + s^{\log r/T} p(\log s),
\quad p(t+T)=p(t)
\end{align*}
which shows that $r>0$ for $f$ to be well-defined.

We now impose the conditions $f(1)=0$ and $f$ convex to check if a suitable $f$ exists.
Since \(t \mapsto e^t\) is an increasing convex function, it is sufficient to check that $g$ is convex on $\R$ to guarantee that $f$ is convex on $(0,\infty)$.
From $f(1)=0$ we obtain $p(0) = -\frac{C}{1-r}$.

As $C/(1-r)$ is constant, the convexity of $g$ depends only on $h(t):=r^{t/T}p(t)$.
Convexity of \(h\) requires that for all $x,y\in\R$ and $\ell\in[0,1]$
\begin{align*}
   r^{(\ell x+(1-\ell)y)/T}&p(\ell x+(1-\ell)y)\\
   &=h(\ell x+(1-\ell)y)
   \\
&\le
\ell h(x)+(1-\ell)h(y)\\
&\le
\ell r^{x/T}p(x)
+
(1-\ell)r^{y/T}p(y).
\end{align*}
Dividing both sides by $r^{(\ell x+(1-\ell)y)/T}>0$ we get
\begin{align*}
p(\ell x+(1-\ell)y)
&\le
\ell r^{\frac{(1-\ell)(x-y)}{T}}p(x)
+
(1-\ell)r^{-\frac{\ell(x-y)}{T}}p(y).
\end{align*}

\end{document}